\documentclass[a4paper,11pt]{article}
\usepackage{pos,multicol}

\title{The hybrid nonet of $\pi_1(1600)$ and $\eta_1(1855)$: analysis and predictions for the kaonic members}
\ShortTitle{$\pi_1(1600)$ and its strange siblings}

\author*{Vanamali Shastry}
\affiliation{Institute of Physics, Jan Kochanowski University, ul. Uniwersytecka 7, P-25-406 Kielce, Poland}

\emailAdd{vshastry@ujk.edu.pl}
\abstract{The hybrid mesons form a part of the exotic spectrum of the standard model. The recent observation of the isoscalar hybrid, called the $\eta_1(1855)$, provides an important step towards the completion of the $1^{-+}$ nonet. In the present work, we analyze the masses and two-body decays of the members of this nonet using a model hadronic Lagrangian. The isovector $\pi_1(1600)$ has been studied extensively - both experimentally and on lattice. We use the available experimental and lattice data to extract the coupling constants. Using these parameters, we analyze the possible decay channels for the hybrid kaons.% We find that the hybrid kaons have to be at least as broad as the $\pi_1(1600)$.
}

\FullConference{%
  41st International Conference on High Energy physics - ICHEP2022\\
  6-13 July, 2022\\
  Bologna, Italy
}

\begin{document}
\maketitle

\section{Introduction}
The rather intricate nature of the strong interactions leads to a magnificent spectrum of QCD containing bound states and resonances. Particularly, the ``exotic'' part of the spectrum consisting of hybrids ($\bar{q}qg$), multiquark states (tetra/penta/hexaquarks), meson molecules, and so on have been studied extensively by the theoreticians and hunted by the experimentalists alike for decades \cite{Meyer:2015eta}, and with considerable success \cite{Workman:2022ynf}. In this contribution, we present the results of a theoretical study involving the lightest $1^{-+}$ hybrids predicted by the nonrelativistic quark models and observed in experiments.\par
This paper is organized as follows. In the next section, we give an account of the model used for the study and the data used for fitting the parameters. In Sec. \ref{sec:kaons}we present the results for the hybrid kaons and in Sec. \ref{sec:SnC} we summarize the paper.

\section{Model\label{sec:model}}
\subsection{Lagrangian}
To study the decays of the $1^{-+}$ hybrid meson nonet ($\Pi_\mu^{hyb}$), we construct a Lagrangian invariant under the $U(3)$ flavor symmetry, charge conjugation and parity reversal. Experimentally, the $\pi_1(1600)$ has been observed in the $\eta\pi$, $\eta'\pi$, $\rho\pi$, $b_1(1235)\pi$ and $f_1\pi$ channels \cite{Workman:2022ynf}. In addition, lattice study predicts the partial widths for the $K^*K$, $f_1'\pi$, and $\rho\omega$ channels \cite{Woss:2020ayi}. Using these inputs, we write the following Lagrangian.
\begin{align}
    \mathcal{L}&=g_{b_1\pi}^c\text{Tr}\Big[\Pi_\mu^{hyb}[P,B^\mu]\Big]\!+\! g_{b_1\pi}^d\text{Tr}\Big[\Pi_{\mu\nu}^{hyb}[P,B^{\mu\nu}]\Big] -g_{\eta\pi}\eta_0\text{Tr}\Big[\Pi_\mu^{hyb}\partial^\mu P\Big]\nonumber\\ &+g_{\rho\pi}\text{Tr}\Big[\tilde{\Pi}_{\mu\nu}^{hyb}[P,V^{\mu\nu}]\Big]
    + g_{\rho\omega} \text{Tr}\Big[\Pi_\mu^{hyb}\{V^{\mu\nu},V_\nu\}\Big]+ g_{f_1\pi} \text{Tr}\Big[\Pi_\mu^{hyb}\{A^{\mu\nu},\partial_\nu P\}\Big].\label{eq:lagJ1}
\end{align}
The Lagrangian contains the following fields: pseudoscalar ($P$), vector ($V_\mu$), axialvector ($A_\mu$), and the pseudovector ($B_\mu$). The coupling constant for the $X^{th}$ channel is given by $g_X$. We now comment on the origins of the different terms present in the Lagrangian. The first two terms represent the $\pi_1(1600)\to b_1(1235)\pi$ decay. The first term gives the lowest order (local, contact) interactions between the three states involved and the second term represents the next higher order (non-local, derivative) interactions. A quick calculation using the principles of addition of angular momenta shows the presence of two angular momentum channels in this decay, and the large ratio of the amplitudes of these partial waves (as given by the experiments \cite{Baker:2003jh}) indicates significant contribution from the higher partial wave. This necessitates the inclusion of the higher order interactions in the Lagrangian \cite{Shastry:2022mhk}. As seen from Eq. \ref{eq:lagJ1}, this is the only channel to be represented by two types of interactions. This is due to the fact that the ratio of the partial wave amplitudes is known fairly accurately for this channel but not for the others.\par
The third term in the Lagrangian describes the decay of the $\pi_1(1600)$ into $\eta^{(\prime)}\pi$ channels. This term breaks the $U_A(1)$ symmetry but is invariant under the $SU(3)$ flavor symmetry \cite{Eshraim:2020ucw}. According to our model, the hybrids can decay into two pseudoscalars only through the breaking of the axial symmetry in the leading order. The fourth term in the Lagrangian describes the decay of the hybrid into a vector and a pseudoscalar state. The hybrid nonet appears in this term in the form of the dual field strength tensor $\tilde{\Pi}_{\mu\nu}^{hyb}=1/2\epsilon_{\mu\nu\rho\lambda}\Pi^{hyb,\mu\nu}$ where, $\epsilon_{\mu\nu\rho\lambda}$ is the Levi-Civita tensor. The presence of the Levi-Civita tensor implies that this term is related to the axial anomaly.
The fifth and the sixth terms of the Lagrangian arise from the hybrids coupling to two left-chiral and right-chiral fields with vector quantum numbers. This leads to the formation of two decay channels - the vector-vector channel and the axial-vector-axial-vector channel. Since the lightest axial-vector mesons are only slightly lighter than the $\pi_1(1600)$, the latter channel is heavily suppressed. However, redefining one of the axial-vector fields via the relation, $A_\mu \rightarrow A_\mu + Z_\pi w_a \partial_\mu P$ leads to the generation of a second decay channel involving an axial-vector and a pseudoscalar state (see Ref. \cite{Parganlija:2012fy} for a detailed discussion). This provides a possible mechanism for the decay of the $\pi_1(1600)$ to the $f_1(1285)\pi$ states. One should note that, even though the fifth and sixth terms in the Lagrangian are derivative interactions (non-local), they still represent the lowest order of the interactions in the corresponding channels. Thus, according to our model, a three-body system containing any of the following combination of states - $\Pi^{hyb} PP$, $\Pi^{hyb} PV$, $\Pi^{hyb}VV$, and $\Pi^{hyb} AP$ cannot interact via local interactions.\par
\subsection{Fitting and Parameters}
We now detail the procedure used to fit the parameters. We begin by listing all the data that are available for the purpose. The information available from the experiments is scarce. The PDG gives the following data \cite{Workman:2022ynf}:
\begin{multicols}{2}
\begin{itemize}
    \item Mass of $\pi_1(1600)$: $1661^{+15}_{-11}$ MeV.
    \item Decay channels of $\pi_1(1600)$: $\rho\pi$, $b_1(1235)\pi$, $\eta^\prime\pi$, and $f_1(1285)\pi$.
    \item Total width of $\pi_1(1600)$: $240\pm 50$ MeV.
    \item The ratio of the branching ratios of the $\pi_1(1600)\to b_1(1235)\pi$ channel in $\ell=2$ and $\ell=0$ channels: $\frac{BR(\pi_1\to b_1\pi)_{\ell=2}}{BR(\pi_1\to b_1\pi)_{\ell=0}}=0.3\pm0.1$ \cite{Baker:2003jh}.
    \item Ratio of the partial widths of $\pi_1(1600)$ decaying into $f_1(1285)\pi$ and $\eta^\prime\pi$: $\frac{\Gamma_{f_1\pi}}{\Gamma_{\eta^\prime\pi}}=3.80\pm0.78$ \cite{E852:2004gpn}.
    \item Ratio of the partial widths of $\pi_1(1600)$ decaying into $\eta\pi$ and $\eta^\prime\pi$: $\frac{\Gamma_{\eta^\prime\pi}}{\Gamma_{\eta\pi}}=5.54\pm 1.1 ^{+1.8}_{-0.27}$ \cite{Kopf:2020yoa}.
    \item Mass of $\pi_1(1400)$: $1354\pm 25$ MeV.
    \item Decay channels of $\pi_1(1400)$: $\eta\pi$.
    \item Total width of $\pi_1(1400)$: $330\pm35$ MeV.
\end{itemize}
\end{multicols}
A word on the two resonances listed above (and in the PDG) is warranted. The non-relativistic quark models predict the existence of one nonet with quantum numbers $1^{-+}$ in the region below $2$ GeV \cite{Meyer:2015eta}. This is also corroborated by other studies. Thus, the observation of two overlapping resonances leads to a puzzling situation. What is peculiar here is that the two states have been observed in complementary decay channels. One is thus lead to strongly consider the possibility of the two states being the same. In fact, this line of argument has been pursued before on the theoretical front where the influence of final state interactions on the pole position of the state was explored \cite{Bass:2001zs}. Along with the recent theoretical study that points to the presence of only one pole \cite{JPAC:2018zyd}, this seems to be the best explanation of the puzzle prompting us to ignore the lighter state.\par

The recent lattice study by the Hadron Spectrum collaboration provides the following data \cite{Woss:2020ayi}:
\begin{multicols}{2}
\begin{itemize}
    \item Mass of $\pi_1(1600)$: $1564$ MeV
    \item Decay width of $\pi_1(1600)$: $139$-$539$ MeV
\end{itemize}
\end{multicols}
\begin{itemize}
    \item The following ranges for the partial widths of the various decay channels of the $\pi_1(1600)$:

\begin{tabular}{llll}
    $\Gamma_{b_1\pi}=139$-$529$ MeV, & $\Gamma_{K^\ast K}=0$-$2$ MeV, & $\Gamma_{f_1\pi}=0$-$24$ MeV, & $\Gamma_{\eta\pi}=0$-$1$ MeV, \\
    $\Gamma_{\rho\pi}=0$-$20$ MeV, & $\Gamma_{\rho\omega}\le 0.15$ MeV,& $\Gamma_{f_1^\prime\pi}=0$-$2$ MeV, & $\Gamma_{\eta^\prime\pi}=0$-$12$ MeV. 
\end{tabular}
\end{itemize}
The lattice simulations assumed the mass of pion to be $700$ MeV. The above values were extrapolated to the physical pion mass. Hence, we consider the central values of the above ranges to be the values for the partial widths and assume a $50\%$ uncertainty. This set of data is the biggest source of uncertainties in our fit. \par
We finally consider the flavor constraints. From the lattice data above, we see that more than one decay channels can come from the same set of nonets - {\it e.g.,} $\rho\pi$ and $K^*K$, $f_1\pi$ and $f_1^\prime\pi$, $\eta\pi$ and $\eta^\prime\pi$. The decay channels are clearly constrained by the flavor symmetry. We are thus in a position to impose further constraints on the parameters. Below is a list of flavor symmetry constraints:

\begin{table}[h]
\centering
\begin{tabular}{lll}
     $\dfrac{\Gamma_{f_1^\prime\pi}}{\Gamma_{f_1\pi}}=0.051\pm 0.015$, & $\dfrac{\Gamma_{K^*K}}{\Gamma_{\rho\pi}}=0.178\pm 0.053$,~ and & $\dfrac{\Gamma_{\eta^\prime\pi}}{\Gamma_{\eta\pi}}=12.7\pm 3.8$
\end{tabular}
\end{table}
\hspace*{-20pt}where, a $30\%$ uncertainty has been assumed. We make the following observations. 1) Even though the uncertainties assumed above are arbitrary, the goodness of fit (as measured by the change in the $\chi^2$ value) changes appreciably only if they are varied by a large amount. 2) The final value listed above (the ratio of the partial widths of $\pi_1(1600)$ in the $\eta^{(\prime)}\pi$ channels) appears to be in contradiction with the experimental data listed earlier. However, these two values are close by considering a $1\sigma$ accuracy. The values of the parameters obtained through a $\chi^2$-fit can be found in Ref. \cite{Shastry:2022mhk}.
\section{Kaons\label{sec:kaons}}
We now present the main results of the present study, restricting the discussions to the kaons.
The mass of the kaon is decided by the mass of the isovector with an additional contribution coming from the presence of (anti-)strange quark \cite{Eshraim:2020ucw}. This contribution is, in fact, half that of the contribution to the $\bar{s}s$ component of the isoscalar state. In the present approach, with only the mass of one of the isoscalars known, this parameter has to be taken to be dependent on the angle of mixing between the isoscalars. An absence of information on the second isoscalar implies that the mixing angle becomes an arbitrary parameter. This forces us to explore various possibilities. Taking into account these uncertainties, we find that the mass of the kaons must be in the range of $1.7-1.76$ GeV\footnote{For details of the various scenarios, see Ref. \cite{Shastry:2022mhk}.}.\par

The kaons are expected to decay into $K_1(1270/1400)\pi$, $(\rho,\omega,\phi)K$, $\eta^{(\prime)}K$,$(\eta,\pi) K^*$, $(\rho,\omega)K^*$, $a_1(1260),b_1(1235),h_1(1170)K$ channels at the tree-level. The partial widths of the these decay channels of the hybrid kaon (with mass $1761$ MeV) are listed in table \ref{tab:decwidkaon}.
\begin{table}[ht]
    \centering
    {\renewcommand{\arraystretch}{1.}
    \begin{tabular}{|c|c|c|c|c|c|}
    \hline
        Channel & \multicolumn{2}{c|}{Width (MeV)} & Channel & \multicolumn{2}{c|}{Width (MeV)}\\\cline{2-3}\cline{5-6}
        & Set$-1$ & Set$-2$ & & Set$-1$ & Set$-2$\\\hline
        $\Gamma_{K_1(1270)\pi}$ & $125\pm 42$ & $48\pm 25$ & $\Gamma_{\rho K}$ & $2.18\pm 0.56$ & $2.19\pm0.57$\\
        $\Gamma_{K_1(1400)\pi}$ & $103\pm 45$ & $98\pm 43$ & $\Gamma_{\omega K}$ & $0.82\pm 0.21$ & $0.82\pm0.21$\\
        $\Gamma_{h_1(1170)K}$ & $1.53\pm 0.28$ & $1.37\pm 0.24$ & $\Gamma_{\phi K}$ & $0.49\pm 0.12$ & $0.49\pm0.13$\\
        $\Gamma_{\eta K}$ & $0.29\pm 0.07$ & $0.29\pm0.07$ & $\Gamma_{K^*\pi}$ & $0.67\pm 0.17$ & $0.67\pm0.17$\\\
        $\Gamma_{\eta^\prime K}$ & $2.77\pm 0.62$ & $2.81\pm0.62$ & $\Gamma_{K^*\eta}$ & $0.30\pm 0.08$ & $0.30\pm0.08$\\
        $\Gamma_{\rho K^*}$ & $0.045\pm 0.016$ & $0.047\pm 0.016$ & $\Gamma_{\omega K^*}$ & $0.011\pm 0.004$ & $0.012\pm 0.004$ \\
        $\Gamma_{a_1 K}$ & $11.0\pm 2.32$ & $11.3\pm 2.35$ & $\Gamma_{b_1 K}$ & $64\pm 14$ & $3.11\pm 2.88$ \\\hline
        & & & $\Gamma_\text{tot}$ & $312\pm 97$ & $170\pm 65$\\\hline
    \end{tabular}}
    \caption{The partial widths and branching ratios of various decay channels and the total width for the hybrid kaon $K_1^{hyb}(1750)$.}
    \label{tab:decwidkaon}
\end{table}
The hybrid kaon is expected to decay predominantly into the axial-kaonic states. The difference in the sign of the $D/S$-ratio influences the partial widths of the $K_1(1270/1400)\pi$ and $b_1(1235)K$ channels significantly. The axial-vector kaons arise out of mixing between the flavor states belonging to the $1^{++}$ and $1^{+-}$ nonets (with mixing angle $\theta_K=56^\circ$ \cite{Divotgey:2013jba}). In addition to this, the $K_1(1400)\pi$ channel closer to the threshold, leading to a smaller $3$-momentum of the decay products. This implies that the contributions of the higher partial wave is smaller for the $K_1(1400)\pi$ decay channel compared to the $K_1(1270)\pi$ channel. Hence, the influence of the sign of the $D/S$-ratio is larger on the $K_1(1270)\pi$ channel than on the $K_1(1400)\pi$ channel. However, it should be noted that the full picture can emerge only when the ratios of the partial wave amplitudes are known for all the decay channels.\par
Of particular interest are the pseudoscalar decay channels - $\eta^{(\prime)}K$. On the theoretical front, these channels arise purely due to the axial anomaly terms in the Lagrangian and hence can be of use in studying the breaking of $U(1)_A$ symmetry and its role in hadron spectrum. On the experimental front, the pseudoscalar decay channels are accessible in multiple processes like the decays of heavy quarkonia ({\it e.g.,} \cite{BESIII:2019dme}), open charm ({\it e.g.,} \cite{Belle:2020fbd}) or open bottom mesons ({\it e.g.,} \cite{BaBar:2009cun}), or eventually in the kaon scattering experiments by COMPASS and/or AMBER (see, \cite{Wallner:2022cxv}).

\section{Summary and Conclusions\label{sec:SnC}}
The light hybrid $1^{-+}$ nonet has been described using a model Lagrangian invariant under the $SU(3)$ flavor symmetry and parity and charge conjugation invariance. The parameters were fitted to the available data and the resulting partial widths for the decays of the $\pi_1(1600)$ match well with the values given by lattice studies. We have also provided predictions for the masses, and partial widths of the hybrid kaons. According to our estimates, the hybrid kaons should be observable in the pseudoscalar decay channels which are already accessible in various experiments.

\section*{Acknowledgement}
This work was performed in collaboration with Christian S. Fischer (Justus Liebig University, Giessen) and Francesco Giacosa (Jan Kochanowski University in Kielce, Poland and Goethe University, Frankfurt).
The author acknowledges financial support from the Polish National Science Centre (NCN) via the OPUS project 2019/ 33/B/ST2/00613.

\end{document}